\documentclass[journal=langd5,manuscript=article,usetitle=true]{achemso}
\usepackage[version=3]{mhchem} 

\usepackage{amssymb,amsfonts,amsmath}
\usepackage{graphicx}
\usepackage{dcolumn}
\usepackage{bm}
\usepackage{subfigure}
\usepackage{color}
\usepackage{amsmath}
\usepackage[normalem]{ulem}
\usepackage[linktocpage,colorlinks=true,linkcolor=blue,citecolor=blue]{hyperref}

\author{Sumesh P. Thampi}
\affiliation{Engineering Mechanics Unit, Jawaharlal Nehru Centre for Advanced Scientific Research, Bangalore 560064, India}
\author{Ronojoy Adhikari}
\affiliation{The Institute of Mathematical Sciences, CIT Campus, Chennai 600113, India}
\author{Rama Govindarajan}
\email{rama@tifrh.res.in}
\affiliation{Engineering Mechanics Unit, Jawaharlal Nehru Centre for Advanced Scientific Research, Bangalore 560064, India}
\title
{Do liquid drops roll or slide on inclined surfaces?}

\begin{document}

\begin{abstract}
We study the motion of a two-dimensional droplet on an inclined surface, under the action of gravity, using a diffuse interface model which allows for arbitrary equilibrium contact angles. The kinematics of motion is analysed by decomposing the gradient of the velocity inside the droplet into a shear and a residual flow. This decomposition helps in distinguishing sliding versus rolling motion of the drop. Our detailed study confirms intuition, in that rolling motion  dominates as the droplet shape approaches a circle, and the viscosity contrast between the droplet and the ambient fluid becomes large. As a consequence of kinematics, the amount of rotation in a general droplet shape follows a universal curve characterised by geometry, and independent of Bond number, surface inclination and equilibrium contact angle, but determined by the slip length and viscosity contrast. Our results open the way towards a rational design of droplet-surface properties, both when rolling motion is desirable (as in self-cleaning hydrophobic droplets) or when it must be prevented (as in insecticide sprays on leaves).
\end{abstract}

\section{Introduction}

When it rains, water drops may be seen moving down glass panes and other solid surfaces. The motion of such a drop is very different from that of a solid object of the same shape and size. On a solid surface inclined to the horizontal, a liquid drop can adapt its shape in response to the forces, and choose between sliding and rolling, or any combination of the two.
The main question we pose here is, what determines this choice? The answer, we show, is that the proportion of roll to slide is determined by just one quantity, for given viscosities and slip length. This quantity is a shape parameter, which measures how far away from circular the drop cross section is. The shape parameter $q$ is in turn determined by the capillary and gravitational forces, as we shall see. Surprisingly, drops of quite different shapes, but with the same $q$ display the same amount of roll.

Experiments do not provide a complete answer so far to the slide or roll question. A complete rolling motion of the drop was observed on superhydrophobic surfaces \cite{quere_2004, quere_2005}. In fact the locus of a fluid element identified with a particle inside the rolling drop was found to be very close to a cycloid. An anomalous increase in the speed of smaller drops was also observed \cite{quere_1999}. This was explained based on the scaling arguments in the model of \citeauthor{mahadevan_1999} wherein the Huygens like motion of non-wetting and very small, viscous drops was considered. The stronger dependency of size on viscous losses compared to the (driving) body forces results in smaller drops moving with very high velocities. They also showed a size independent velocity of large pancake drops on non-wetting surfaces. In contrast, some studies \cite{Sakai_2006, Sakai_2008} observed purely slipping motion of water drops on superhydrophobic surfaces, which could be distinguished from a partial slipping observed on normal hydrophobic surfaces. Here the slip velocity at the solid surface was directly measured using imaging techniques, and the discrepancy between this and the total forward velocity of the drop, if any, was attributed to rolling motion. Experimental investigations on hydrophilic surfaces have focused less on this question, but have mostly concentrated instead on instabilities associated with the receding front \cite{limat_2005, limat_2005b}. A noticeable feature of experiments is that a vast majority of them have studied the limiting cases of perfect wetting and zero wetting. Relatively little has been done on the more frequently observed case, where the surface is between completely hydrophobic and hydrophilic, and so the contact angle lies between $0^o$ and $180^o$. Our study addresses the entire range of contact angles.

A majority of theoretical studies too have addressed the hydrophobic or hydrophilic limits, where simplifications are possible. When the contact angle is small, the lubrication approximation of the Navier-Stokes equation is a good model \cite{dussan_1983}, to describe the dynamics, with the additional assumption of a parabolic velocity profile \cite{limat_2005, limat_2005b}. The other limit, when the contact angle approaches $180 ^{\circ}$, has been analysed \cite{mahadevan_1999}, who provide scaling estimates for drop velocity. For contact angles in between these limits, techniques such as molecular dynamics \cite{Muller_2008} and lattice Boltzmann simulations \cite{Moradi_2011} have been employed. Such studies have concentrated on the net velocity of the droplet and its dependence on the driving force and the contact angle, rather than a detailed analysis of the flow field. A splitting of the velocity field, obtained using a phase field model, into slip and roll was done recently \cite{yeomans_2010}. This study was conducted for nearly circular drops of sizes smaller than the capillary length. In this size range it is possible to estimate the quantum of roll using the behaviour at the center of the drop. However, for larger drops, the shape can deviate from circular by a large amount, and the amount of roll has to be obtained differently, as we will see. Contact line motion, which is imperative for determining droplet motion, is not well understood \cite{eggers_2009}. An approximation frequently used in this context is the prescription of rolling motion near the contact line \cite{Clarke_1995, Rame_1996}. Our approach is to take into consideration rolling motion everywhere in the bulk. The association between the two is a measure of nonlocal hydrodynamic effects of contact line movement \cite{ShikhmurzaevBook}.

The intermediate case is of interest in several applications. A liquid drop that rolls down a plane inclined to the horizontal, and picks up tiny solid particles as it does so, is crucial in some self-cleaning devices, but a drop that stays stationary is needed in spray and coating applications \cite{eggers_2009, Book_deGennes, rothstein_2010}. Besides this, just the fact that the downward motion of a liquid drop is so much more complex than that of a solid object begs better understanding. Even when the drop is static, the reaction forces and moments provided by the supporting surface are qualitatively different from those for a solid body \cite{sumesh_2011}, and the complexity for a moving drop is compounded by viscous and local inertial stresses. Our study is not limited to low inertia.

Our flow situation is that of a liquid drop positioned on a solid surface inclined to gravity at an angle $\alpha$. Around the liquid drop is a second fluid, lighter and less viscous than the liquid making up the drop. The break-up of the motion into slide and roll is done in a non-standard way, to ensure that the vorticity which contributes to shearing motion alone is left out of the reckoning for measuring the roll. We do this because the standard decomposition of velocity gradient tensor \cite{batchelorbook} is insufficient to measure rolling motion, since the antisymmetric part of the velocity gradient tensor contains contributions to both shear and roll. Ours is a hybrid numerical approach combining lattice Boltzmann and a diffuse interface model for binary fluids. The model allows for different viscosities for the two fluids and a large range of contact angles on smooth surfaces. We analyse the entire spectrum of shapes, for a range of the relevant non-dimensional parameters including sizes larger than capillary length. Non-intuitively, it will be seen that the rolling behaviour can be universally described by a shape parameter, independent of capillary and gravitational forces. For small Bond numbers, it is possible to predict this shape parameter from the static shapes. We will also see that reducing the viscosity of the external fluid increases the rotation inside the drop, surprisingly, with out much change in the drop shape. The role of slip length at the contact line is also investigated in determining the amount of roll.

\section{Theory and Simulations}

While a complete analytical solution of Navier - Stokes and Cahn - Hilliard equations may be difficult, it is possible to derive asymptotic and scaling solutions, as often done in fluid mechanics, to provide valuable insight. As discussed earlier, an example of this in the small hydrophobic drop context has already been given by \citeauthor{mahadevan_1999}. We choose to perform numerical solutions, which too pose considerable challenges, due to the multiple length scales present, and the coupling between the evolving field and the interface shape \cite{zaleski1998}. We verify our results through scaling analysis. We use a coupled system of equations describing the dynamics of a conserved order parameter $\psi$, defined as a normalised density difference, and the conserved momentum density $\rho {\bf u}$, where $\rho$ and ${\bf u}$ are the total density and the local fluid velocity \cite{Kendon2001, anderson1998}. The order parameter dynamics is described by a Cahn-Hilliard equation (CHE), which includes advection by fluid flow and relaxation due to chemical potential gradients \cite{ChaikinBook}. This is coupled to a Navier-Stokes equation (NSE) \cite{batchelorbook, LandauBook} with additional stress densities arising from the order parameter.
\begin{align}
 \partial_t \psi + \nabla \cdotp \left( \mathbf{u}\psi\right) &= \nabla \cdotp \left( M\nabla \mu \right)\label{eqn:che}\\
 \partial_t (\rho \mathbf{u}) + \nabla \cdotp (\rho \mathbf{uu}) &= -\nabla p + \eta \nabla^2 \mathbf{u} + \psi \nabla \mu + \mathbf{G}
\label{eqn:chens}
\end{align}
together with the continuity equation for the density. In the above, $p$ stands for the pressure,  $\eta$ is the shear viscosity and $\mathbf{G}$ is the gravitational force density. The mobility $M$ is the constant of proportionality in the linear phenomenological law relating the flux of $\psi$ to the thermodynamic force $\nabla \mu$.

The equilibrium thermodynamics of the fluid is described by the Landau free-energy functional \cite{ChaikinBook, RowlinsonBook}
\begin{equation}
F(\psi) = \int(\frac{A}{2}\psi^2 + \frac{B}{4} \psi^4+\frac{K}{2}\left|\nabla\psi|^2\right)d\mathbf{r},
\label{eqn:Feng}
\end{equation}
with $A<0$, $B>0$ and $\mathbf{r}$ stands for the spatial dimensions. This functional form of free energy provides two uniform solutions $\psi = \pm \sqrt{A/B}$ coexisting across a fluid interface representing the drop and the surrounding fluid. The three parameters $A$, $B$, and $K$ control the interfacial thickness $ \xi = \sqrt{2K/A}$ and the interfacial energy $ \gamma = \frac{2}{3} \sqrt{2KA^3/B^2}$ \cite{Kendon2001}. Gradients in the order parameter produce additional stresses, $\psi\nabla\mu = \nabla \cdot \boldsymbol{\sigma}^{\psi}$, which includes Laplace and Marangoni stresses due to a fluid-fluid interface \cite{anderson1998}. Here $\mu = A \psi + B \psi^3 - K \nabla^2 \psi$ is the chemical potential.

We use a hybrid algorithm by combining the lattice Boltzmann (LB) method for hydrodynamics and method of lines for the order parameter dynamics \cite{sumeshlb_2011}. Force densities such as the divergences of order parameter stresses and gravity \cite{greated_2000} are included in the modified LB method used here \cite{nash2008}. We implement the body force only on one fluid, which is equivalent to solving the NSE with the Boussinesq approximation. We use a $D3Q15$ model and collision integral is a single relaxation time ($\tau$) approximation. In order to ensure that we are in the incompressible limit, we must have $|\mathbf{G}z| << \rho c_s^2$ where $z$ is the vertical extent of the simulation domain, which means that thermodynamic pressure is large compared to the hydrostatic pressure difference. The viscosity is obtained as $\eta = \tau c_s^2$ where $c_s = 1/\sqrt{3}$ is the sound speed in LB units. The spatial discretization of the CHE is based on a finite-volume formulation \cite{RotenBerg2008} and this  set of equations is temporally integrated using a Runge-Kutta algorithm.  The interested reader is referred to \cite{sumeshlb_2011} for a detailed description of this method.

The wall is placed at the $\frac{1}{2}$ grid point, as is usual in the bounce back schemes used to represent wall in LB method \cite{SucciBook}. Also an additional free energy functional of the form $\frac{C}{2} \psi_s^2 + H \psi_s$, where $\psi_s$ is the value of order parameter at the wall, is introduced at the wall boundaries to model solid-liquid surface tension \cite{Desplat2001, yeomanscont_2002}. Minimisation of the energy functional near the wall gives $C \psi_s + H = K \nabla \psi \cdot \mathbf{n}$ where $\mathbf{n}$ is normal to the wall and acts as the boundary condition for $\psi$. By tuning the parameters $C$ and $H$ we can modify the wetting properties of the surface. Thus, when the fluid - fluid interface meets the solid surface a contact line is formed with an equilibrium contact angle, $\theta_e$. This is same as Young's equilibrium contact angle describing the force balance at the contact line at equilibrium indicating the wetting properties of the surface and can be related to the model parameters. It is found sufficient to retain only the linear term of the surface energy functional \cite{Desplat2001, yeomanscont_2002} and thus desired contact angles are obtained. Since very small and very large contact angles are not captured accurately in this model, we restrict our simulations as far as possible to the wide range of intermediate contact angles.

Simulations have been performed in a box of dimensions $512 \times 256 \times 1$ LB units for a two dimensional drop. Wall boundary conditions are applied on two sides and periodic boundary conditions are applied on the other two sides. The simulation is initiated with a semicircular drop of radius $L = 60$ sitting on one wall, which is inclined at an angle $\alpha$ to the horizontal. In response to gravity ($\mathbf{G}$) and surface forces, the drop starts moving on the solid surface. We impose a smooth surface, i.e., one which does not display any hysteresis. The simulation is continued till the drop reaches a steady state velocity $\mathbf{V}$. We define the Bond number as $Bo \equiv L^2|\mathbf{G}|/\sigma$, the Reynolds number as $Re \equiv L V \rho/\eta$ and the Capillary number as $Ca \equiv \eta V/\sigma$ where $\eta$ and $\sigma$ are viscosity and surface tension respectively. The thick interface and the contact line region are excluded in quantitative estimates of roll and slide.

As discussed above, it is necessary to split the flow field into shear, slide and roll. Now vorticity is a local quantity which includes both the solid body rotation and the shearing motion of a fluid element. It thus cannot distinguish between them. We therefore make a triple decomposition of the velocity gradient tensor into a straining part, a simple shear flow part, and a rigid body rotation. The shear part is removed from the total vorticity to obtain a residual vorticity characterising the local rotation. This method was introduced by \citeauthor{kolar_2007} in a different context. Given a velocity field, $\mathbf{u}$, in the $x-y$ plane the velocity gradient tensor is a $2 \times 2$~matrix. The symmetric part of the velocity gradient tensor has eigen values  $\pm s/2$ where $s=\sqrt{4u_x^2+(u_y+v_x)^2}$. This is the strain rate tensor in the principal coordinates and represents the total straining of the fluid element. The antisymmetric part is characterised by the vorticity $\omega = v_x-u_y$. The residual vorticity and strain may be written respectively as
\begin{eqnarray}
 \omega_{res} &=& 0 \quad \text{if} \quad |s|\ge|\omega| \nonumber\\
              & = & \text{sgn}(\omega) \left[|\omega|-|s|\right] \quad \text{if} \quad |s|\le|\omega|,\\
s_{res}&=& \text{sgn}(s)\left[|s|-|\omega|\right] \quad \text{if} \quad |s| \ge |\omega|\nonumber \\
&=& 0 \quad \text{if} \quad |s| \le |\omega|.
\end{eqnarray}
More details about the triple decomposition of the velocity gradient tensor may be found in \cite{kolar_2007} and explained in detail in the SI. A pictorial representation of the method is provided in Fig. 3 in the SI.

\begin{figure*}
\centering
\includegraphics[width=\linewidth]{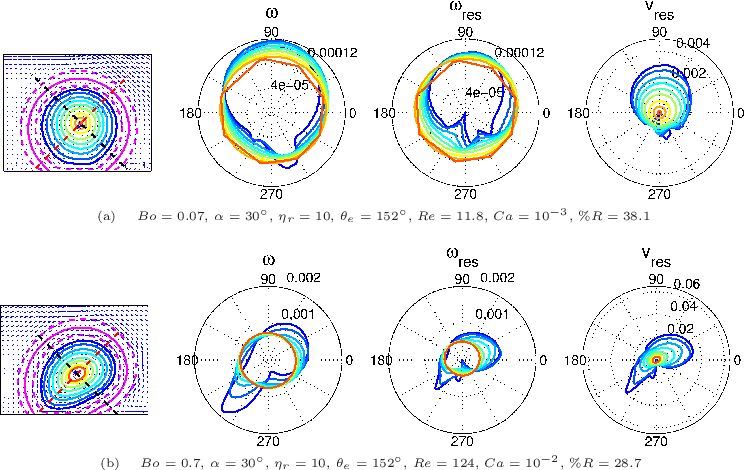}
\caption{Effect of gravity on dynamics, with gravity being higher by a factor of $10$ in figure (b) than in (a). The drop shape and streamline patterns are seen in the left-most panel. In the next three panels respectively, the vorticity ($\omega$), the residual vorticity ($\omega_{res}$) and the residual angular velocity ($v_{res}$) are illustrated, by polar plots, in a coordinate frame moving with the center of mass of the drop. In these polar plots, the color of each curve matches with the corresponding streamline in the left-most figure. The azimuthal angle is measured from a line parallel to the solid plate, and the radial location indicates the magnitude of the quantity being plotted. In this and subsequent figures, the thick interface denoted by the magenta dashed lines spans from $\psi=0.9$ to $\psi=-0.9$. The drop is moving on a surface inclined to the horizontal, and the black dashed line indicates the direction of gravity. The red dashed line is normal to gravity.}
\label{fig:geodef}
\end{figure*}

If a solid body is rolling on an inclined surface with an angular velocity of $N$, then the corresponding vorticity is $2N$. Therefore, we can find the average residual vorticity inside a drop and calculate a corresponding forward velocity of the drop corresponding to the roll as 
\begin{equation}
 V_{rolling} = \frac{\text{Average}(\omega_{res})}{2} \times \frac{\text{Height of the drop}}{2}.
\end{equation}
Here we take the radius of the drop to be half of the maximum height. Then a quantity called percentage rotation, denoted by $\%R$, is calculated based on the total translational velocity $V$ of the drop, as 
\begin{equation}
\%R=V_{rolling}/V \times 100.
\label{perr}
\end{equation}
In Fig. \ref{fig:geodef} one may see that increasing gravity by a factor of $10$ increases the translational velocity, and therefore the Re and Ca, by an order of magnitude. This standard scaling \cite{Kang_2001} holds throughout our simulations (see Fig. 8 in SI) in contrast to \cite{mahadevan_1999} where the two limiting cases, namely the Huygens like motion of small, non-wetting viscous drops and the size independent motion of pancake drops are described. Though the terminal drop velocity increases linearly with driving force the associated deformation reduces the percentage rotation $\%R$, which is by 10\% in the specific case of Fig. \ref{fig:geodef}. This estimate of \%R is different from the roll versus slip velocity as defined in \cite{yeomans_2010}, who do not distinguish the shear vorticity from residual vorticity. This would be a good estimate for a drop which is practically circular, as is the case in that study, but not for a deformed drop.

\begin{figure*}
\includegraphics[width=\linewidth]{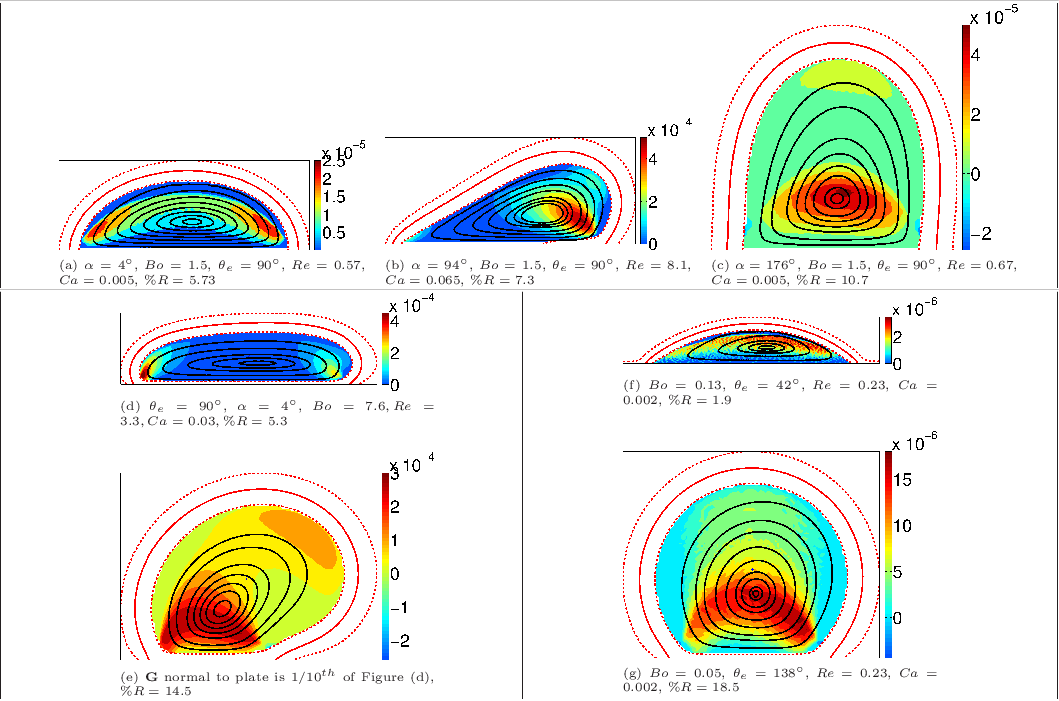}
\caption{Effect of drop shape on the rolling behavior of drops. The color represents the residual vorticity, while the continuous lines are streamlines. In Fig. \ref{fig:shapedep}a, \ref{fig:shapedep}b and \ref{fig:shapedep}c the inclination of the solid plate is progressively increased, maintaining all other parameters the same. The pendant drop in \ref{fig:shapedep}c, with a height comparable to its radius displays increased solid body rotation. The effect of normal component of gravity is illustrated in Fig. \ref{fig:shapedep}d and \ref{fig:shapedep}e. In these two figures, the only difference is that the normal component of gravity is artificially suppressed to $1/10^{th}$ of its value in \ref{fig:shapedep}e. The tangential component being the same, the drops display a comparable forward velocity. The effect of $\theta_e$ on the shape and rolling behavior is shown in Fig. \ref{fig:shapedep}f and \ref{fig:shapedep}g. Here, $Re$ and $Ca$ are maintained the same by adjusting the $Bo$. The $\% R$ is larger when shape is closer to a circle in all cases.}
\label{fig:shapedep}
\end{figure*}

\section{Results and discussion}

\subsection{Drop shapes and rolling dynamics}

We will first look at the effect of gravitational and capillary forces. The effect of increasing Bond number is illustrated in Fig. \ref{fig:geodef}. A larger driving force means larger deformation, so the drop deviates further from its equilibrium shape at zero plate inclination. The streamlines are plotted in the center of mass frame. Fixing the center at the innermost streamline, the vorticity and residual vorticity along different streamlines are plotted as functions of azimuthal angle, thus mapping the entire vorticity field inside the drop. Finally the angular velocity based on residual vorticity, $v_{res} = r\omega_{res}$ where $r$ is taken as the radial distance from the center of the innermost streamline, is plotted as a function of azimuthal angle. A perfect solid body rotation, such as that of a solid wheel would have appeared as concentric circles in this plot. The drop is more circular at low gravity (small Bond number), and the angular velocity for different streamlines is similar, except for a slight loss of symmetry in that the outer streamlines move faster at the top and slower at the bottom. This means that the dynamics is closer to that of a solid object at low gravity, which is evidenced also by the residual vorticities for different streamlines being closer to each other than in the high gravity case. It is seen that even at low gravity, evaluating only the total vorticity would render several features of the dynamics invisible to us. For example, in a region near the rear contact line, the last two panel show that shear vorticity dominates, breaking the up-down symmetry.  

In the case of large $Bo$, the drop is longer normal to gravity, with a clear breakdown in left-right symmetry in its shape, as illustrated in Fig. \ref{fig:geodef}b. Except for the very center of the drop there is no resemblance to solid body rotation or even to tank-treading. This is reflected in the angular velocity plot as well. The residual vorticity is higher in the direction of elongation. Interestingly the residual vorticity is now higher near the rear of the drop, exactly where it was lower at low $Bo$. This is because at higher gravity the rear of the drop has a tendency to lift off the surface. A given fluid element accelerates and decelerates significantly as it moves on a streamline. 

Unlike the experiments of drops on superhydrophobic surfaces \cite{quere_2004, quere_2005} and the analysis of limiting cases in \cite{mahadevan_1999},  the present study is valid over a wide range of parameters, and in particular for a range of non-spherical shapes, surface wettability and gravity. For small droplets, the crucial assumption in \cite{mahadevan_1999} was that the deviation of the shape from a sphere is very small. Relevant length scales of the  deformation as a response to gravity were thence derived. These scaling arguments break down when $\theta_e \ne 180^{\circ}$ due to a finite contact area as we have in our simulations. Also, since we do not restrict our analysis to small $Bo$, the changes in the surface energy need not scale with that of gravitational potential energy and hence the scaling relations of \cite{mahadevan_1999} will not be valid here.

\begin{figure}
 \includegraphics[trim =0 0 2 2, clip, width=0.8\linewidth]{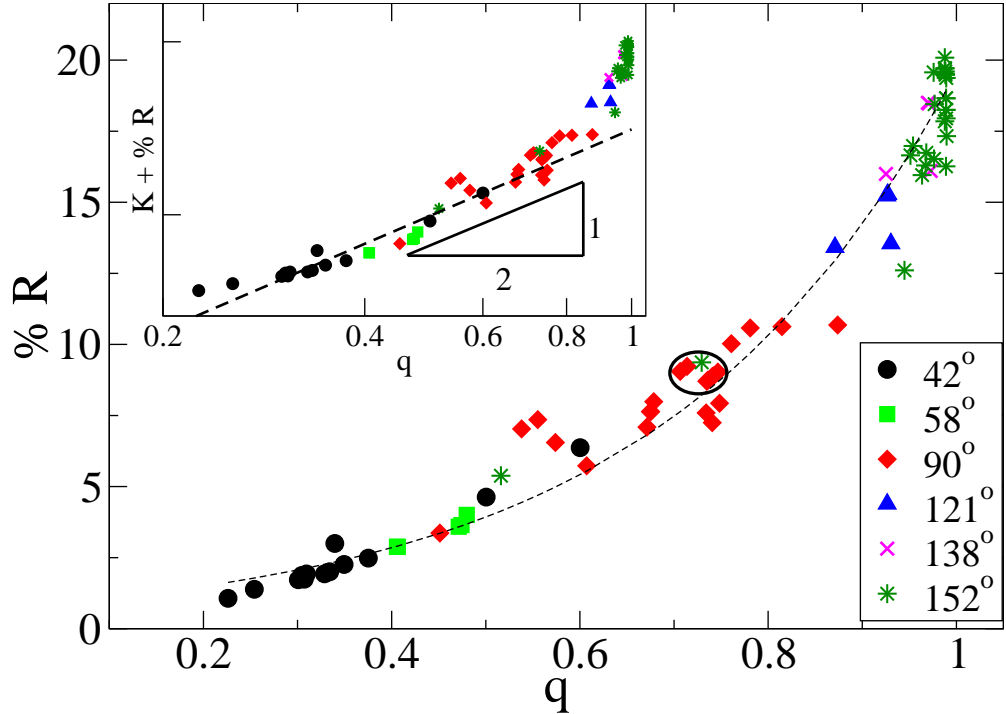}
\caption{\%R as a function of isoperimetric quotient is plotted for different sets of simulations. Each symbol represents a particular $\theta_e$. Within each set, $\alpha$ varies from $4^{\circ}$ to $176^{\circ}$ and $Bo$ ranges from $5 \times 10^{-3}$ to $1.5$. \%R of drops for a wide variation in parameters fall on this curve. An exponential curve fitted through all data points is also shown. This is a universal curve giving percentage rotation as a function of shape factor for a given slip length and viscosity ratio. In the inset, the same data is plotted in log-log scale to show the consistency with scaling arguments (see text). Here, the constant, $K$, is chosen as 10. 
 Details of the data points circled are shown in Fig. \ref{fig:diffshape}.}
\label{fig:qall}
\end{figure}

\begin{figure*}
\centering
\includegraphics[width=\linewidth]{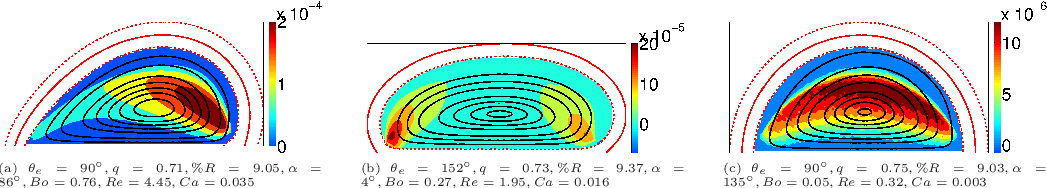}
\caption{Three different drop shapes and the corresponding streamline and residual vorticity patterns. The shapes are distinct from each other, but the isoperimetric quotient is approximately the same. This results in the same roll fraction. These cases lie within the circle marked in Fig. \ref{fig:qall}.}
\label{fig:diffshape}
\end{figure*}

Hence, one may infer that an important property that determines the motion of a drop is its geometry. Needless to say, the geometry is determined by the various forces. A variety of shapes of drops in steady motion along the plate are illustrated in Fig. \ref{fig:shapedep}.
We first discuss the effect of the tilt angle of the plate, which determines the ratio of the components of gravity normal and tangential to the plate. The normal component alone does not produce any movement of the drop, but both components contribute to deciding the shape, and thence the dynamics. As illustrated in Fig. \ref{fig:shapedep}a, \ref{fig:shapedep}b, \ref{fig:shapedep}c the drop tends to lift off from the plate and display an increased height as tilt increases. In turn the percentage rotation increases, and is seen to be highest for a tilt angle of $176^{\circ}$. This means that a pendant liquid drop is much more likely to roll than a sessile one which is the same in all other ways. The presence of corners and deformed parts of the drop always increases the shear vorticity locally. As the plate inclination changes, not only does the ratio of normal to tangential forces change, but also each of the magnitudes. In order to study the effect of the ratio alone, Fig. \ref{fig:shapedep}d and \ref{fig:shapedep}e are made varying the normal force component, while artificially maintaining the tangential force the same. This corresponds to a simultaneous variation in plate inclination and gravity to achieve the same settling velocity. One can clearly see that as the normal component of gravity is reduced, the shape becomes more and more elongated in the direction normal to the plate and this increases the amount of rotation considerably. In Fig. \ref{fig:shapedep}f and \ref{fig:shapedep}g, the effect of another important parameter, the equilibrium contact angle, on the drop shape and hence on the dynamics, is illustrated. Here gravity is adjusted so that the drop attains the same terminal settling velocity and hence the same $Re$ in both cases, ensuring that the effects of inertia are nullified in this comparison. In contrast to the case of $\theta_e = 138^{\circ}$, one may see that no rotation is present when $\theta_e = 42^{\circ}$. Here the entire vorticity of the fluid elements in the latter case can be attributed to that associated with shear. Such drops may be dealt with the lubrication approximation \cite{dussan_1983}. As the equilibrium contact angle increases, the percentage rotation increases, with a maximum in the case of an almost circular drop.

Hence we conclude that the deviation from a circular shape plays an important role in determining the dynamics. One can then suitably define a shape parameter to describe the closeness of the shape to a circle, for example, the isoperimetric quotient
\begin{equation}
 q = \frac{4 \pi \times \text{Area}}{\text{Perimeter}^2}.
\end{equation}
This ratio is unity for a circle and is less than this value for any other shape, since a circle has the least circumference for a given area.

Fig. \ref{fig:qall} shows that for a fixed slip length and viscosity ratio, a universal relationship between the percentage rotation and isoperimetric quotient is obeyed. As expected, the percentage rotation is higher for a shape which is closer to a circle. It is however of interest to note that irrespective of the parameters such as Bond number, plate inclination and equilibrium contact angle which determines the shape and the deformation, the percentage of roll is a function \textit{only} of this quantity. The collapse of data from a large number of simulations spanning a wide range of $\theta_e$, $\alpha$ and $Bo$ in Fig. \ref{fig:qall} is a strong indication of this. However the viscosity, the mobility and the viscosity ratio are kept fixed in the simulations shown so far. Having these parameters the same, any change in capillary or gravitational forces will change the shape of the drop and then the amount of rotation can be uniquely provided by Fig. \ref{fig:qall}. Though the dependence of the shape is intuitive, the unique dependence on one parameter, and the functional form, are not. The dependence may be shown to be a consequence of kinematics as discussed below.

Taking cues from the velocity gradient decomposition, we may write the total velocity gradient, $\nabla u = \nabla u_{slip}+\nabla u_{roll}$, as that resulting from two types of motion. Then, in an order of magnitude estimate, $\nabla u \sim U/ \sqrt{A}$ where $\sqrt{A}$ is a characteristic size of the drop, $A$ being the area. The deviation of the drop from a circular shape may be represented by defining different length measures $h$ and $l$, normal to, and along the solid plate, respectively. Thus the shape factor $q \sim A/(h+l)^2$. Given that $l \sim A/h$, 
$q=q(h/\sqrt{A})$. Since the drop is in motion only along the direction on the solid plate, a good order of magnitude estimate of the shear force, as in \cite{mahadevan_1999}, is just $\mu U/h$, so $\nabla u_{slip} \sim O(U/h)$. Now $\nabla u_{roll}$ is nothing but the residual vorticity, or the effective angular velocity, of the drop. Thus, using Eq. \ref{perr} and multiplying throughout by $\sqrt{A}/U$, we obtain that $\%R \sim \textnormal{function}(q) - K$ where $K$ is a constant coming out of exact numerical values. We thus show that not only does the fraction of roll versus slide depend only on geometry, this dependence on geometry comes in only in the form of the shape factor $q$. It may be noted $\sqrt{q} \sim h/\sqrt{A}$ for small enough $q$. As illustrated in the inset of Fig. \ref{fig:qall},  $\%R \sim \sqrt{q}$ for a range of $q$, consistent with the scaling arguments. This square root dependence is only for small $q$, since at large $q$, we notice that the fraction of roll increases more rapidly with $q$ than a square root behaviour would predict. It is interesting to note that capillary, gravitational or wetting parameters do not appear explicitly in determining \%R, consistent with our observations.
We wish to highlight Fig. \ref{fig:diffshape} where results from three different simulations, circled in Fig. \ref{fig:qall}, are shown. Here the shapes are distinct from each other, but the isoperimetric quotient is approximately the same. This results in the same roll fraction in all these cases. 

\begin{figure*}
\includegraphics[trim=0 0 0 0, clip, width=\linewidth]{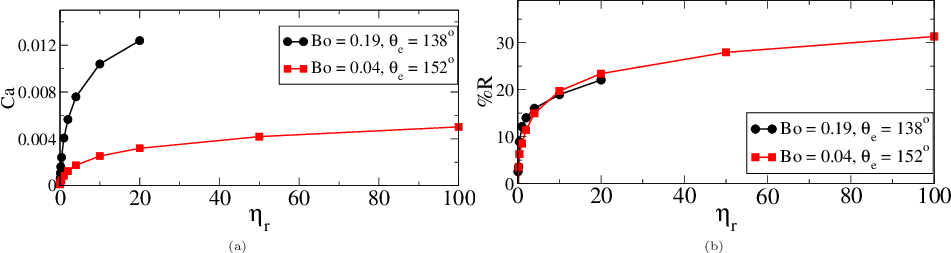}
\caption{Effect of external fluid viscosity on the $Ca$ and the rolling behavior is illustrated. As the viscosity ratio $\eta_r$ increases, which corresponds to a reduction in the viscosity of the external fluid, the drop translates faster, but rolling motion inside the drop increases. Low viscosity drops in a higher viscosity fluid are seen to almost slide on the wall rather than execute a rolling motion.}
\label{fig:muratio}
\end{figure*}

\begin{figure*}
\includegraphics[trim = 0 0 0 0, clip, width=\linewidth]{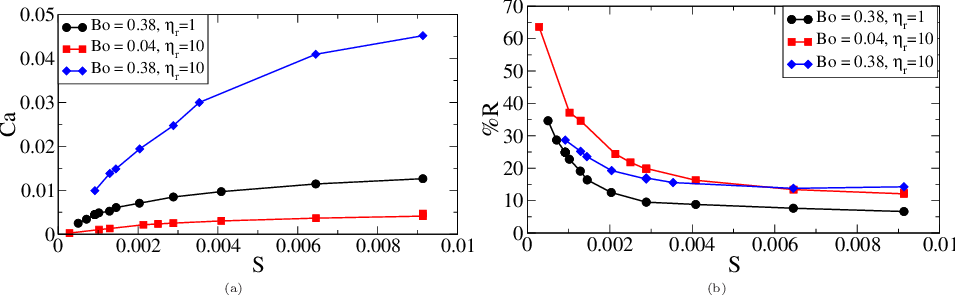}
\caption{$Ca$ and rolling behavior are plotted as a function of the nondimensional slip length, $S$. In order to obtain a range of $S$, the viscosity was independently varied by three orders of magnitude and mobility by one order of magnitude. The equilibrium contact angle is $152^{\circ}$. Larger slip length at the contact line results in larger translational velocity of the drop. Percentage rotation, \%R, also strongly depends on the slip length.}
\label{fig:slips}
\end{figure*}

\subsection{Effect of viscosity}

When the viscosity of the external fluid is reduced, the settling velocity and hence the $Ca$, as expected, increase as shown in Fig. \ref{fig:muratio}. Here $\eta_r$ is defined as the ratio of viscosity of the drop to that of the external fluid. Also the percentage of rolling motion is larger. In line with this, one may expect a significant amount of rolling in case of a water-air system where viscosity contrast is large. As the viscosity of the external fluid increases beyond that of the drop, the dynamics shifts towards the external fluid. The drops slide in that case. This too is consistent with intuition, since a `bubble' will simply slide in a liquid rather than roll when moving on a surface. In these cases, one may observe that, despite the geometry remaining similar, the percentage rotation increases when the viscosity ratio increases. Therefore the universal curve describing the dependence on isoperimetric quotient will be shifted appropriately by a change in the viscosity ratio
. However, when the viscosity ratio is fixed it is found that the magnitude of the viscosity does not explicitly affect the share of rolling motion (SI Text).

\subsection{Effect of slip length}
\label{sec:slip}

Apart from the solid body rotation, which gives a forward velocity to the entire drop, the contact line moves due to the slip provided by the diffusion of the order parameter \cite{jacqmin_2000}. Thus the inescapable contact line singularity arising from the classic no-slip boundary condition is relieved at the contact line in the diffuse interface models. In the neighbourhood of contact line strong gradients in the order parameter generates diffusion. As described Eq. \ref{eqn:che}, then strong flows are generated which advects the order parameter and thus generating the contact line slip\cite{seppecher_1996}. Balancing the advection and diffusion of the order parameter across the interface provides a length scale for this process as $\lambda = \sqrt{\eta M}$. Though there exist debates on the exact length scale and we refer the reader to \citeauthor{feng_2010} for details, it has been shown that this slip length is the same as that used in the slip-induced movement of the contact line in sharp interface models \cite{cox_1986, cox_1998}. Hence this slip length is not an artificial parameter \cite{feng_2010} and is not an unexpected result \cite{dussan_1976}. Therefore we can use $\lambda$ as a measure of slip at the contact line. This means that either mobility or viscosity can be independently or simultaneously varied to change the slip at the contact line. Slip length is here defined using the viscosity of the drop. We define a nondimensional slip length as $S = \lambda/L $. 

Both viscosity and mobility are independently varied by at least one order of magnitude in Fig. \ref{fig:slips} to obtain a range of $S$. As the slip length increases, the $Ca$ also increases as illustrated in Fig. \ref{fig:slips}a. Intuitively, a slipping drop on an inclined surface will roll less. This is verified in our simulations as shown in Fig. \ref{fig:slips}b wherein the importance of slip length in determining the amount of rotation inside the drop may be inferred. And this dependence appears to be exponential. These simulations have been done for a drop on a hydrophobic surface ($\theta_e = 152^{\circ}$) and for relatively small $Bo$ numbers. Other parameters such as capillary and gravitation forces (in effect the shape) and viscosity ratio will also play an important role in deciding the $\%R$. Therefore a combined phase space spanning Fig. \ref{fig:qall}, \ref{fig:muratio} \& \ref{fig:slips} should be considered when comparing with experiments. Larger percentage rotations than those shown, which would correspond to smaller slip lengths could not be obtained reliably with the present numerical simulations. Also, we have not considered cases where the drops will lift off from the surface or break up due to strong shears generated.

In our simulations $S$ varies from $10^{-3}$ to $10^{-2}$. In the light of experimental evidence where slip length varies from $nm$ to $\mu m$ \cite{meinhart_2002, lohse_2009, vinogradova_2009, rothstein_2010} and since we concentrate on the bulk motion of fluid elements, we expect that our simulations are relevant in several practical applications independent of the particular mechanism responsible for the slip. It is possible to incorporate slip into the standard scaling relations, vis. $Bo \sim Ca$, thus making the effect of slip length explicit in the studies, as exemplified in SI Text.

\section{Conclusions}

To conclude, a universal curve is obtained for the motion of a droplet
on a surface inclined at an angle to gravity. This curve describes the
dependence of the fraction of rolling versus sliding motion, as a
function of a single quantity, namely the isoperimetric quotient $q$,
for a given viscosity ratio and slip length. This dependence is termed
universal, since it is obeyed irrespective of the Bond number, plate
inclination and equilibrium contact angle. Remarkably, drops of widely
different shapes but the same $q$ display the same amount of roll.
These results are obtained by a hybrid simulation method implementing
lattice Boltzmann algorithm with diffuse interface model and have been justified based on scaling arguments. A residual
vorticity is introduced, and computed from the velocity field inside
the drop. This, rather than the commonly used total vorticity \cite{yeomans_2010}, is a
better measure of characterizing the global rolling motion of the
drop. The external fluid certainly affects the drop motion, allowing a
larger degree of rolling motion when its viscosity is small compared
to the viscosity of the drop. Surprisingly this happens without much
change in the drop shape. Similarly the importance of the slip
mechanism of the contact line is discussed in relation to the rolling
motion inside the drop.

This study answers the question of roll versus slip in a drop on an
inclined surface. Though the analysis is done in two dimensions it can
be easily extended to three dimensions and the observations are easily
verifiable from experiments. The method of characterizing rotation
from residual vorticity may be applied to velocity fields obtained
through PIV measurements and correlated to a shape parameter defined
in 3D. We expect our main conclusions such as the unique dependence of amount of roll on shape, viscosity of liquids and the slip length to remain valid in 3D as well. Further, since this work gives detailed predictions about bulk
motion, the effect of slip length can also be easily verified from
experiments on patterned surfaces. Any correlation so obtained would be valuable since this analysis relates a microscopic length scale to the macroscopic dynamics. There is a general paucity of
experimental investigations of binary fluids in the presence of a
wall. We hope that our results will motivate readers to perform
experiments to verify our findings and deepen insight into the
dynamics of drops on inclined surfaces.

\acknowledgement

We thank an anonymous referee for an important suggestion. We gratefully acknowledge Ignacio Pagonabarraga for the fruitful discussions.

\suppinfo

A supporting document containing several details of simulation method and results is available. This information is available free of charge via the Internet at http://pubs.acs.org/.
\bibliography{reference}

\begin{figure*}[h]
\centering
\includegraphics[width=\linewidth]{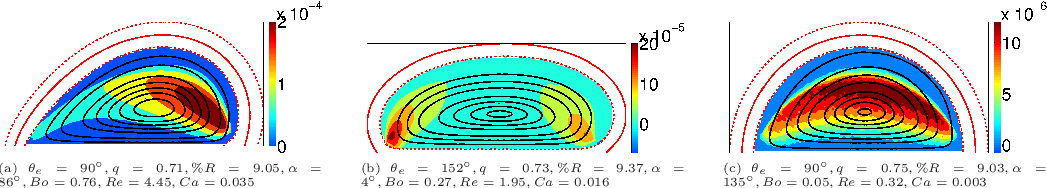}
TOC Figure: Three different drop shapes and the corresponding streamline and residual vorticity patterns. The shapes are distinct from each other, but the isoperimetric quotient is approximately the same. This results in the same roll fraction.
\end{figure*}

\end{document}